\documentclass[fleqn,usenatbib]{mnras}

\usepackage{newtxtext,newtxmath}

\usepackage[T1]{fontenc}
\usepackage{ae,aecompl}
\usepackage{bm,breqn}
\usepackage[normalem]{ulem}

\usepackage{graphicx}	
\usepackage{amsmath}	
\usepackage{amssymb}	
\usepackage{cleveref}
\usepackage{color}

\crefformat{section}{\S#2#1#3} 
\crefformat{subsection}{\S#2#1#3}
\crefformat{subsubsection}{\S#2#1#3}

\newcommand{\planck}{{\it Planck}}

\title[Reconstructing Probability Distributions with GPs]{Reconstructing Probability Distributions with Gaussian Processes}

\author[T. McClintock]{
Thomas McClintock,$^{1}$\thanks{E-mail: mcclintock@bnl.gov} Eduardo Rozo,$^{2}$
\\
$^{1}$Brookhaven National Laboratory, Physics Department, Building 510, Upton, New York, 11973\\
$^{2}$Department of Physics, University of Arizona, Tuscon, AZ 85721, USA
}

\date{Accepted XXX. Received YYY; in original form ZZZ}

\pubyear{2019}

\begin{document}
\label{firstpage}
\pagerange{\pageref{firstpage}--\pageref{lastpage}}
\maketitle

\begin{abstract}
Modern cosmological analyses constrain physical parameters using Markov Chain Monte Carlo (MCMC) or similar sampling techniques. Oftentimes, these techniques are computationally expensive to run and require up to thousands of CPU hours to complete. Here we present a method for reconstructing the log-probability distributions of completed experiments from an existing MCMC chain (or any set of posterior samples). The reconstruction is performed using Gaussian process regression for interpolating the log-probability. This allows for easy resampling, importance sampling, marginalization, testing different samplers, investigating chain convergence, and other operations. As an example use-case, we reconstruct the posterior distribution of the most recent \planck\ 2018 analysis.  We then resample the posterior, and generate a new MCMC chain with forty times as many points in only thirty minutes. Our likelihood reconstruction tool can be found online at \url{https://github.com/tmcclintock/AReconstructionTool}.
\end{abstract}



\section{Introduction}

Cosmology has entered the ``precision era'' where experiments aim to put ever-tighter constraints on a small number of physical parameters that describe our Universe. Since the discovery of the accelerated expansion of the universe using Type Ia supernovae \citep{Riess1998_SNIa,Perlmutter1999_SNIa} up through recent analyses of the large scale structure of the universe \citep[e.g.][]{6dF2011_Cosmo,BOSS2017_Cosmo,DES2018_KP3x2,Planck2018_Cosmology,HSC2019_Cosmo,KiDSVik450_2019}, the primary means of making such constraints has stayed the same. Using Markov Chain Monte Carlo (MCMC) or similar techniques, most analyses end up producing samples of the posterior probability distribution of the parameters of interest given the data under some model. These samples are used to compute statistical moments of the marginalized distributions of the parameters, which are reported as final results.

The plethora of cosmic probes means that one can combine experiments in order to enhance overall constraining power \citep[][for recent examples]{DES2018_H0BBN,Baxter2019_5x2}. Unfortunately, the technical and detailed nature of individual analyses can make it difficult, if not impossible, to make certain combinations. Indeed, even reproducing past results can be laborious. So much so, that a cottage industry has sprouted just to create software for performing joint analyses and enabling reproducibility (e.g. Cosmosis\footnote{\url{https://bitbucket.org/joezuntz/cosmosis/wiki/Home}} \citealt{Zuntz2015_Cosmosis}, Cobaya\footnote{\url{https://github.com/CobayaSampler/cobaya}}, CosmoMC\footnote{\url{https://cosmologist.info/cosmomc/}} \citealt{Lewis2002_CosmoMC}, and MontePython\footnote{\url{https://github.com/brinckmann/montepython_public}} \citealt{Audren2013_MontePython}). As indicated by the breadth of detailed instructions for installing and configuring these packages, resampling a posterior from a completed analysis is not as simple as pushing a button. Compounding this problem is the shear computational power required. For example, an MCMC chain from the DES Year 1 key project combining cosmic shear, galaxy-galaxy lensing, and galaxy clustering took approximately three thousand total CPU hours. Combining DES with an analysis that has a different computational bottleneck would exacerbate the problem, potentially making a full joint analysis prohibitively expensive. A final hurdle to resampling is the simple fact that analysis code is often not released until long after chains have been made public. For instance, the posterior chains from the \citet{Planck2018_Cosmology} analysis have been released, but software for evaluating their likelihood has not. Nevertheless, performing joint analyses is essential for discovering new physics. Thus, we seek a solution to the related problems of reproducibility, expense, and inconvenience.

In an effort to tackle these problems simultaneously, we propose using Gaussian process regression to reconstruct the posterior probability distribution of an experiment. Gaussian processes are probabilistic interpolators that can be used to model complicated functions in many dimensions. They have been successfully used in the past in cosmology for emulation \citep{Lawrence2017_MT2,DarkQuest1_2018,Zhai2018,McClintock2019_HMFEmu}, and have many applications in machine learning \citep{RasmussenWilliamsGPs}. Our method allows for reconstructing and resampling probability distributions given a small set of samples taken from an MCMC chain. Sampling from the reconstructed distribution can be orders of magnitude faster than evaluating the likelihood of a given experiment. For instance, large scale structure likelihoods are bottlenecked by Boltzmann codes, which do not need to be evaluated using our technique.

As a demonstration, we take the MCMC chains released in \citet{Planck2018_Cosmology} and reconstruct the (unreleased) posterior probability of that analysis. By sampling from our reconstructed distribution, we recover constraints on all free parameters. Our MCMC chain has forty times as many samples even after accounting for burn-in, and can be produced in minutes on a laptop computer.

The paper is structured as follows: \cref{sec:gpr} covers the basics of Gaussian process regression. In \cref{sec:distribution_modeling_with_GPR} we discuss how to select training points from an MCMC chain and how we construct the Gaussian process for interpolation. \cref{sec:basic_examples} shows some examples of our method, including reconstructing the \citet{Planck2018_Cosmology} posterior. Finally, \cref{sec:conclusions} concludes with a discussion of some applications of our technique as well as potential extensions.


\section{Regression with Gaussian Processes} \label{sec:gpr}

Here we discuss the basics of regression with Gaussian processes. Gaussian processes have many uses and a thorough discussion of the topic is provided in \citet{RasmussenWilliamsGPs}. In this work, we use the Gaussian process implementation found in the \textsc{Python} package \texttt{george}\footnote{\url{http://george.readthedocs.io/en/latest}} \citep{Ambikasaran2015_George}.

In our case, we are have $N$ samples of the log-probability of some distribution $\ln\mathcal{P}$ at locations in parameter space given by $\theta$. In this section, we will write $\ln\mathcal{P}=y$, so that our training data takes the form of a series of $N$ distinct points in parameter space along with the corresponding log-probability, i.e.  $\{\bm{\theta}_i,y_i\}_{i=1}^N$. We know the values of $y$ exactly, since they are taken directly from an MCMC chain or an equivalent set of samples as the result of an experiment.

We want to build an interpolator in order to be able to predict the log-probability of a new location in parameter space $y^*(\theta^*)$ with no assumptions about the specific function form of the log-probability. Gaussian processes are a tool that is ideally suited for this task. Gaussian processes assume that the covariance between two samples $y_i(\bm{\theta}_i)$ and $y_j(\bm{\theta}_j)$ depends only on the values of $\bm{\theta}_i$ and $\bm{\theta}_j$, and that together $\{\bm{\theta},y\}$ form a multivariate normal distribution. Our samples $\{\bm{\theta}_i,y_i\}_{i=1}^N$ represent $N$ draws from this distribution. Therefore, a prediction for $y^*(\bm{\theta}^*)$ is the mean of the conditional multivariate normal given our samples according to
\begin{equation}
    \label{eq:gp_mean_y}
    \langle y^*|\theta^*,\{\bm{\theta}_i,y_i\}_{i=1}^N\rangle = \mu(\bm{\theta}^*) + \Sigma_*^T\Sigma^{-1} \mathbf{y}\,.
\end{equation}
In \autoref{eq:gp_mean_y}, $\mu(\bm{\theta}^*)$ is the so-called mean function, and $\mathbf{y}$ is a column vector containing $\{y_i\}_{i=1}^N$. $\Sigma_*^T$ denotes the $1\times N$ row vector of the covariance evaluated at all pairs of $\bm{\theta}^*$ with $\{\bm{\theta}_i\}_{i=1}^N$, while $\Sigma$ is the $N \times N$ matrix containing the covariance between all pairs of $\{\bm{\theta}_i\}_{i=1}^N$. Conventionally, the mean function is set to 0 assuming that the training data has been standardized such that the sample mean is $\frac{1}{N}\sum y_i = 0$, although it need not be. This mean function can also be understood as the estimate for $y^*$ when $\bm{\theta}^*$ is far away from the known samples, or the value that the Gaussian process extrapolates to. Since we are building an interpolator for log-probabilities, we require this mean function to be small. We set it to a value less than the minimum log-probability in our samples, $\mu < \min(\{y_i\})$, and assume that our samples represent the ``region of interest'' that the original MCMC chain derives from. That is to say, if the MCMC chain does not well sample the parameter space, one cannot expect the interpolator to do any better. We found that the exact value of $\mu$ does not affect our results, as long as its absolute value is not too large compared to the training points. In other words, if $|\mu| \gg \max(\{|y_i|\})$ the interpolator will not perform well.

The elements of the covariance matrices are given by the kernel function $[\Sigma]_{ij} = k(\bm{\theta}_i,\bm{\theta}_j)$. Specifying this kernel function completes the Gaussian process. We found that the squared exponential kernel function is sufficient. This kernel is given by
\begin{equation}
    \label{eq:sq_exp_kernel}
    k(\bm{\theta}_i,\bm{\theta}_j) = \exp\left[ - \frac{(\bm{\theta}_i - \bm{\theta}_j)^2}{2l^2}\right]\,.
\end{equation}
\autoref{eq:sq_exp_kernel} states that the covariance between two points in parameter space follows a Gaussian, so that as points become farther separated they influence each other less. This range of influence is controlled by the hyperparameter $l$, which is the characteristic length scale of correlations within the data. The value of $l$ is found by maximizing the likelihood
\begin{dmath}
    \label{eq:gp_likelihood}
\ln\mathcal{L}(\mathbf{y}|l,\bm{\theta}) =
    - \frac{1}{2}\left[(\mathbf{y}-\mu)^T\Sigma^{-1}(l,\bm{\theta})(\mathbf{y}-\mu) 
    + \log\det(\Sigma(l,\bm{\theta})) 
    + d\log 2\pi\right]\,. 
\end{dmath}
The left term is the $\chi^2$ of the data fit by the model while the middle term penalizes excessively large covariance. The right most term is a normalization constant that depends on the dimensionality $d$ of the data (i.e. $\bm{\theta}$ is a vector in a $d$-dimensional parameter space). Upon maximizing \autoref{eq:gp_likelihood}, $l$ is fixed and we can predict new values of the log-probability according to \autoref{eq:gp_mean_y}.

Notably, we use only a single hyperparameter in \autoref{eq:sq_exp_kernel}. Usually, when using a Gaussian process to interpolate in many dimensions, the characteristic length scale in each of those dimensions can be different, and one requires a different $l$ for each dimension. For instance, the characteristic length scale for interpolating over $\Omega_m$ should be very different than the length scale required to interpolate over $A_s$, since these parameters differ by many orders of magnitude. In the following section we detail how we transform the training samples in our model so that we require only a single hyperparameter.


\section{Modeling Probability Distributions with Gaussian Process Regression} \label{sec:distribution_modeling_with_GPR}

MCMC chains represent probability distributions via the number density of points from the chain at any given location in parameter space. However, samplers also record the log-probability of each entry in the chain, although this quantity is oftentimes not used. In contrast, we reconstruct the log-probability as a function of the parameters in the chain while only using a special selection of samples. In \cref{sec:selecting_training_samples}, we detail our method of selecting training points from an MCMC chain. These training points are then used to optimize a Gaussian process. Next, in \cref{sec:performing_regression} we detail how we construct the interpolator and perform regression.


\subsection{Selecting Training Samples} \label{sec:selecting_training_samples}

Each sample in an MCMC chain is a vector in a $d$-dimensional parameter space $\mathbf{p} = \{p_i\}_{i=1}^{d}$. The sample mean and sample covariance matrix of the parameters can be computed from the chain according to
\begin{align}
    \hat{\bar{p}}_i &= \frac{1}{N_c} \sum_{k=1}^{N_c} p_{i,k} \\
    \hat{\mathbf{C}}(p_i,p_j) &= \frac{1}{N_c}\sum_{k=1}^{N_c}(\hat{\bar{p}}_i - p_{i,k})(\hat{\bar{p}}_j - p_{j,k})\,,
\end{align}
where there are $N_c$ samples in the chain indexed by $k$, $\hat{\bar{p}}_i$ is the sample mean of the $i$-th parameter, and $\hat{\mathbf{C}}$ is a $d\times d$ dimensional matrix. We eigendecompose $\hat{\mathbf{C}}$ into components
\begin{equation}
    \hat{\mathbf{C}} = \mathbf{Q}\mathbf{\Lambda}\mathbf{Q}^T\,,
\end{equation}
where $\mathbf{\Lambda}$ is a diagonal matrix containing the eigenvalues, and $\mathbf{Q}$ is an orthogonal matrix with columns containing the eigenvectors. $\mathbf{Q}$ can be understood as a rotation matrix that rotates any sample in parameter space into a space in which the parameters are all orthogonal.

Next, we select $N_s$ samples in a $d$-dimensional unit hypercube, $\mathbf{x}' = \{x_i'\}_{i=1}^d$ centered on the origin. Our results are not strongly sensitive to how these samples are selected. By default, we use a simple Latin-Hypercube sampling as implemented in the package \textsc{pyDOE}\footnote{\url{https://github.com/clicumu/pyDOE2 }}\citep{Surowiec2017_pyDOE}. It is unlikely that this is an optimal choice, and selecting training samples for Gaussian processes remains an open topic of research \citep{Heitmann2009_Coyote2,DarkQuest1_2018}. Since our results are not sensitive to this choice of sampling, we defer further study of this topic to a future work.

\begin{figure}
	\begin{center}
	\includegraphics[width=\linewidth]{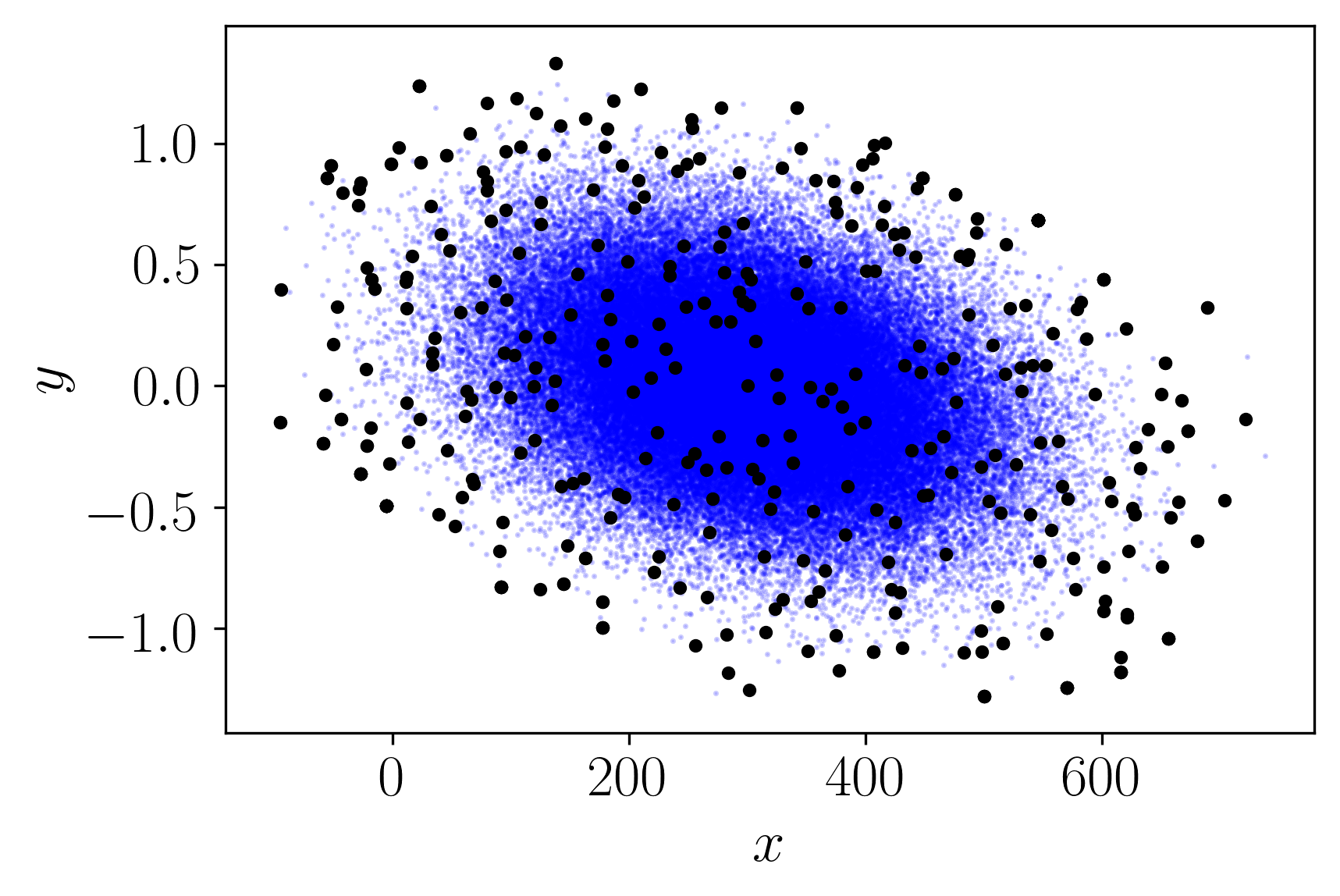}
	\caption{Black points are select locations from the underlying MCMC chain (light blue) that are designed to span the probability surface. Subsampling the chain is necessary since Gaussian processes do not scale well with a large number of samples.
	\label{fig:training_points}}
	\end{center}
\end{figure}

The $N_s$ samples are then rotated into parameter space, scaled by a constant factor $s$, and mean centered according to
\begin{equation}
    \label{eq:LH_to_p_transformation}
    \mathbf{x} = s\mathbf{\Lambda}^{\frac{1}{2}}\mathbf{Q}^T\mathbf{x}' + \hat{\bar{\mathbf{p}}}\,,
\end{equation}
where $\hat{\bar{\mathbf{p}}}$ is a vector containing the sample mean of the parameters in the chain, and $\mathbf{\Lambda}^{\frac{1}{2}}$ is a diagonal matrix containing the square roots of the eigenvalues. While $\mathbf{x}'$ was a vector in a unit hypercube, $\mathbf{x}$ is a vector in parameter space. The constant factor $s$ ``spreads out'' the random samples. Without it, the samples are contained approximately within the $1\sigma$ level of the chain. Setting $s\sim8$ results in points reaching the $\sim4\sigma$ level. \autoref{fig:training_points} show an example of a selection of training points from a two parameter MCMC chain that sampled a multivariate Gaussian likelihood. The black points are spread evenly across the sample space, while the blue points of the chain are clustered as one might expect.

For a given $\mathbf{x}\in\{\mathbf{x}_i\}_{i=1}^{N_s}$ the distance between that point and a sample in the chain is given by
\begin{equation}
    d^2 = (\mathbf{x} - \mathbf{p})^T\hat{\mathbf{C}}^{-1}(\mathbf{x} - \mathbf{p})\,.
\end{equation}
Thus, to select our final set of training points $\mathbf{p}_{\mathbf{x}}$ we find the nearest point in the chain for each $\mathbf{x}$ according to
\begin{equation}
    \mathbf{p}_{\mathbf{x}} = \operatorname*{argmin}_{\mathbf{p}} d^2(\mathbf{x},\mathbf{p}) | \forall \mathbf{p} \in \{\mathbf{p}_i\}_{i=1}^{N_c}\,.
\end{equation}
This set of points $\{\mathbf{p}_{\mathbf{x},i}\}_{i=1}^{N_s}$ is a subset of the original MCMC chain that follows the Latin-Hypercube design we chose. There is a chance that multiple $\mathbf{x}$ select the same sample $\mathbf{p}$, however this can be checked and corrected for by sampling additional training points.

This method for selecting training points is well suited for subsampling points from a smooth, nearly Gaussian distribution. If the points in an MCMC chain have a complicated distribution, one may need to tailor the selection of training points to achieve a desired result. Additionally, one may choose specific points from the chain to include in the subsample, such as the maximum a-posteriori probability point (i.e. the most likely point in the chain), or points in the tails.


\subsection{Performing Regression} \label{sec:performing_regression}

Constructing the Gaussian process and performing regression requires additional work. Given a new location in parameter space $\mathbf{p}^*$, simply applying the inverse transformation of \autoref{eq:LH_to_p_transformation} is insufficient. This is because in this basis, the characteristic length in each dimension is different, meaning we require one hyperparameter for each free parameter in the MCMC chain. This becomes infeasible for large parameter spaces such as the 27 dimensional \planck\ analysis.

Instead, we construct the Gaussian process in a special basis. First, we transform each point in the original MCMC chain according to
\begin{equation}
    \mathbf{q}' = \mathbf{Q}\mathbf{p}\,.
\end{equation}
We then compute the sample mean $\hat{\bar{\mathbf{q}}}'$ and sample covariance $\hat{\mathbf{C}}_{\mathbf{q}'}$ for these points. In this basis, the new samples $\{\mathbf{q}_i'\}_{i=1}^{N_c}$ have a diagonal covariance matrix, however each dimension of $\mathbf{q}'$ has a different variance. Thus, our final transformation is to standardize these points by subtracting the sample mean and dividing each dimension by the sample standard deviation
\begin{equation}
    \label{eq:q_final}
    \mathbf{q}_{i} = \frac{\mathbf{q}_{i}' - \hat{\mathbf{q}}'}{\hat{\bm{\sigma}}'}\ \forall \mathbf{q}_i'\in\{\mathbf{q}_i'\}_{i=1}^{N_c}\,,
\end{equation}
where $\hat{\bm{\sigma}}'$ is a vector containing the sample standard deviations for each dimension. In this parameter space, the set of points $\{\mathbf{q}_i\}_{i=1}^{N_c}$ has zero mean, unit variance, and no correlation between the parameters in each dimension.

Each of the training points $\{\mathbf{p}_{\mathbf{x},i}\}_{i=1}^{N_s}$ selected in \cref{sec:selecting_training_samples} correspond to individual $\mathbf{q}$ points in this space. We select this set $\{\mathbf{q}_i\}_{i=1}^{N_s}$ in order to train the Gaussian process. Since each dimension of parameter space in this basis has unit variance, the Gaussian process requires only a single hyperparameter to successfully model the distribution.

When predicting the log-probability at a new point in parameter space $\mathbf{p}^*$, the process is straightforward. First, rotate the parameter vector by multiplying by $\mathbf{Q}$. Second, subtract the sample mean $\hat{\bar{\mathbf{q}}}'$ and divide by the sample standard deviation for each dimension $\hat{\bm{\sigma}}'$. Third, estimate the log-probability according to \autoref{eq:gp_mean_y}. This completes the reconstruction of the probability distribution sampled by the input MCMC chain. We can now use the Gaussian process to resample the distribution in order to perform calculations including marginalization, importance sampling, joint sampling, and computing Bayesian evidence. Select examples are presented in the following section.

\section{Example Use Cases} \label{sec:basic_examples}

The ability to quickly resample a probability distribution has many potential applications. In this section we highlight two examples that have immediate utility.


\subsection{Sampling Disjoint Parameter Spaces} \label{sec:disjoint_distributions}

\begin{figure}
	\begin{center}
	\includegraphics[width=\linewidth]{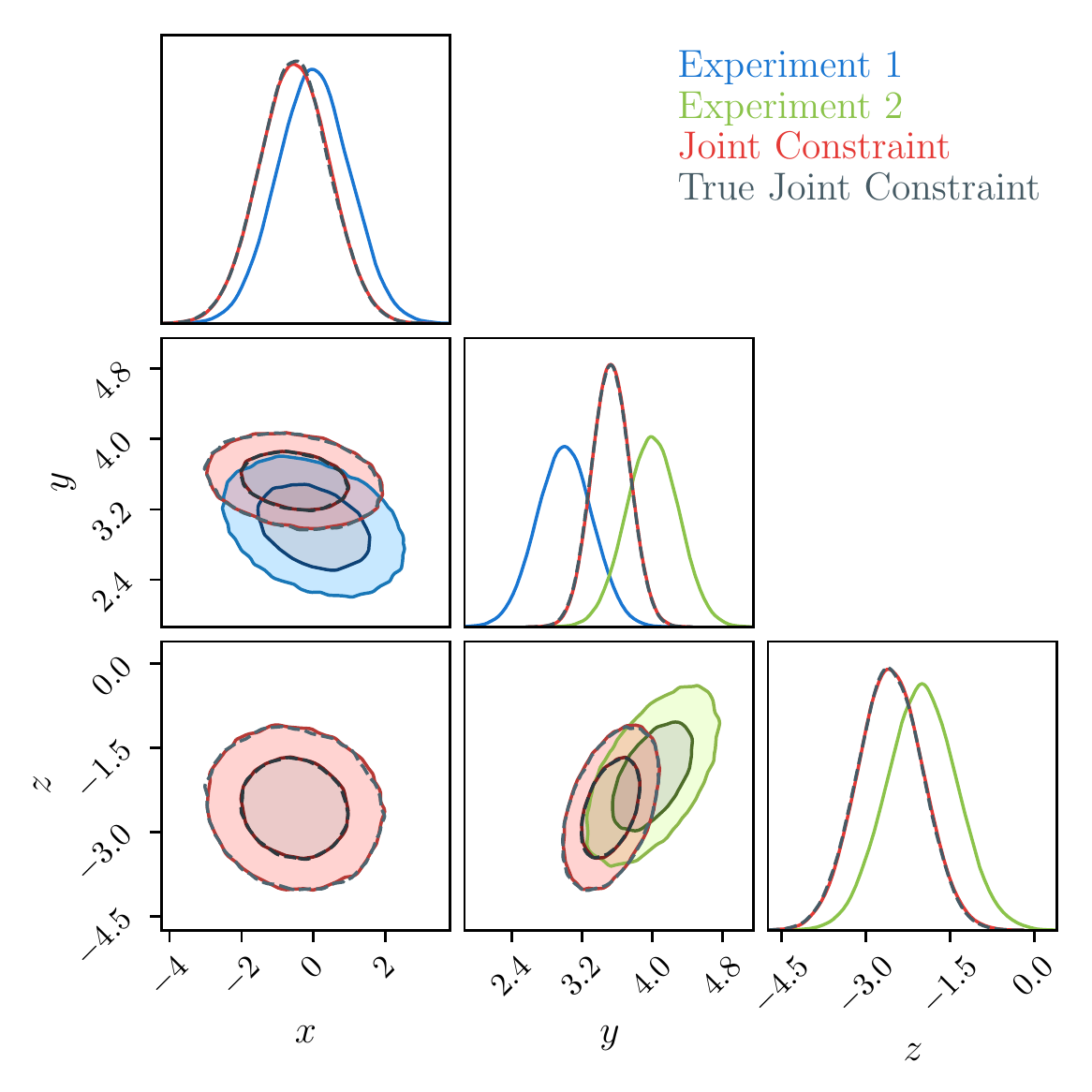}
	\caption{Reconstructing two posterior distributions with disjoint parameter spaces and resampling both to make a joint constraint. The blue contours are the constraints from experiment 1 on parameters $\{x,y\}$, while the green contours show the constraints from experiment 2 on parameters $\{y,z\}$. The (Gaussian) likelihoods of both experiments are reconstructed using our method, and jointly sampled to produce the red contours. The black dashed contours show the true joint constraint, which overlaps exactly with our reconstructed result.
	\label{fig:disjoint_figure}}
	\end{center}
\end{figure}

Oftentimes two experiments share parameters of interest but contain their own set of nuisance parameters. Using our method, we can reconstruct the individual probability distributions of two or more experiments in order to jointly sample all parameters at once. \autoref{fig:disjoint_figure} demonstrates this using a simple example involving three parameters. Suppose we have one experiment that samples $\{x,y\}$ and another that samples $\{y,z\}$, and MCMC samples for both (blue and green contours in the figure, respectively). We reconstruct the probability distributions of both experiments. Assuming that the two experiments are independent and have flat priors, we can simply multiply the two reconstructed probabilities to determine the joint likelihood of the two experiments, leading to the red contours in \autoref{fig:disjoint_figure}. As an aside, we note that if either experiment had informative priors, we would need to remove one set of priors in order to construct a valid posterior. This demonstrates the utility of our method for quickly combining any set of experiments just using their MCMC chains.

\subsection{Resampling the Planck 2018 Posterior} \label{sec:resampling_planck}

As a cosmologically relevant example, we illustrate how our code can be utilized to reconstruct the \citet{Planck2018_Cosmology} TT, TE, EE+lowE posterior.

\begin{figure*}
	\begin{center}
	\includegraphics[width=\linewidth]{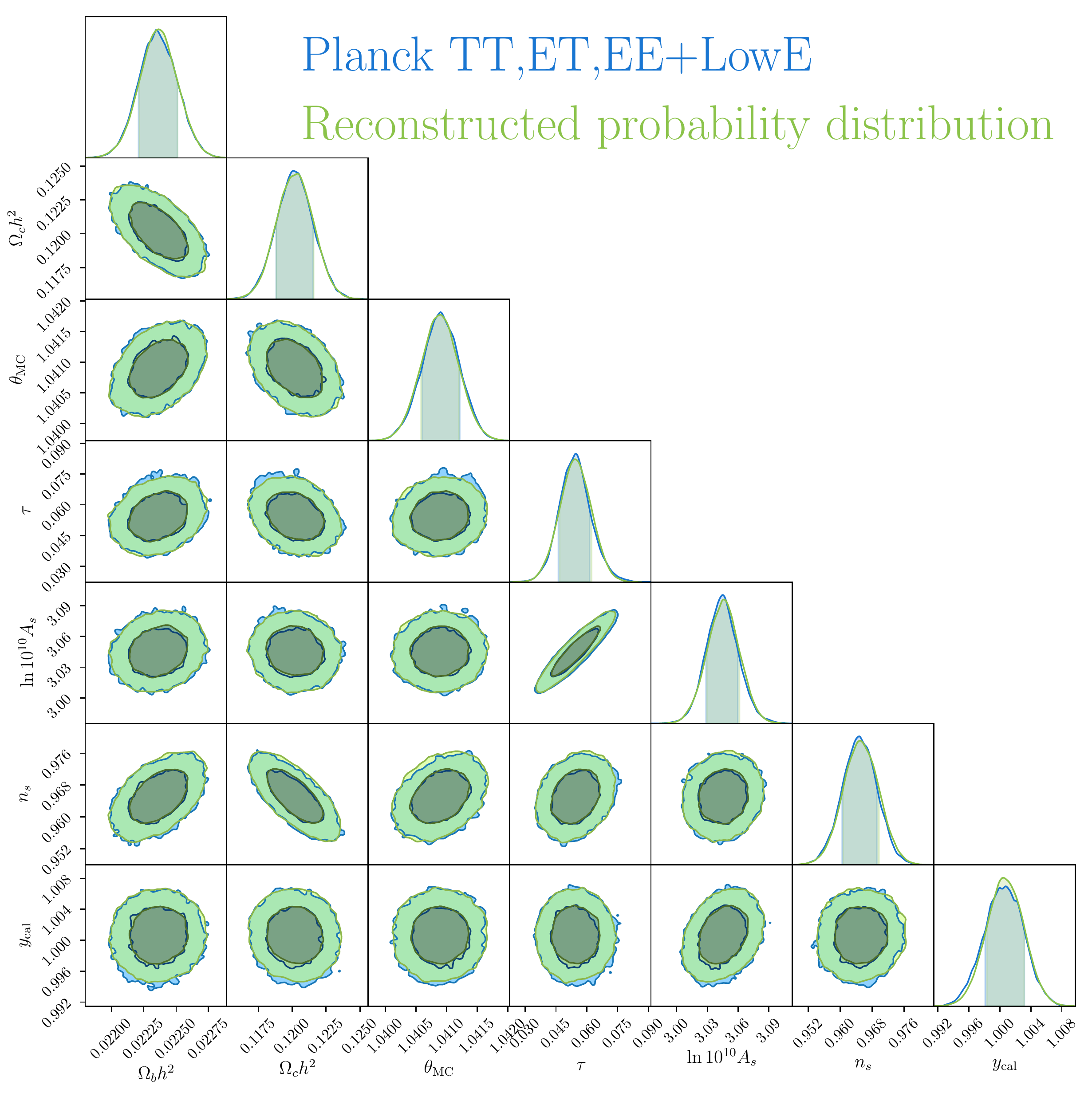}
	\caption{The posterior distribution of the first seven parameters of the \citet{Planck2018_Cosmology} MCMC chains. Blue contours are the original contours without smoothing. Green contours come from sampling our reconstructed posterior probability distribution. We use only 1200 points from the original MCMC chain to reconstruct the posterior. The MCMC chain of the reconstructed distribution is forty times longer than the original chain, even after removing burn-in. The overlap of the posteriors demonstrates our reconstruction tool accurately models the posterior surface, with slight differences appearing due to shot noise in the original chain. \label{fig:planck_figure}}
	\end{center}
\end{figure*}

\autoref{fig:planck_figure} shows constraints on the first seven parameters of the Planck chain in blue. The entire distribution has 27 free parameters, however all parameters past the first six are nuisance parameters. We checked that our reconstruction successfully recovered those 20 nuisance parameters, but present only the cosmologically interesting parameters here. 

Note that the blue contours are less smooth than those found in \citet{Planck2018_Cosmology}. This is because we have not applied a kernel density estimate for smoothing to highlight the ability of our reconstruction to achieve good convergence. The green contours come from sampling the reconstructed probability distribution using our method. This new MCMC chain is forty times longer than the original Planck chain after removing burn-in.

\begin{figure}
	\begin{center}
	\includegraphics[width=\linewidth]{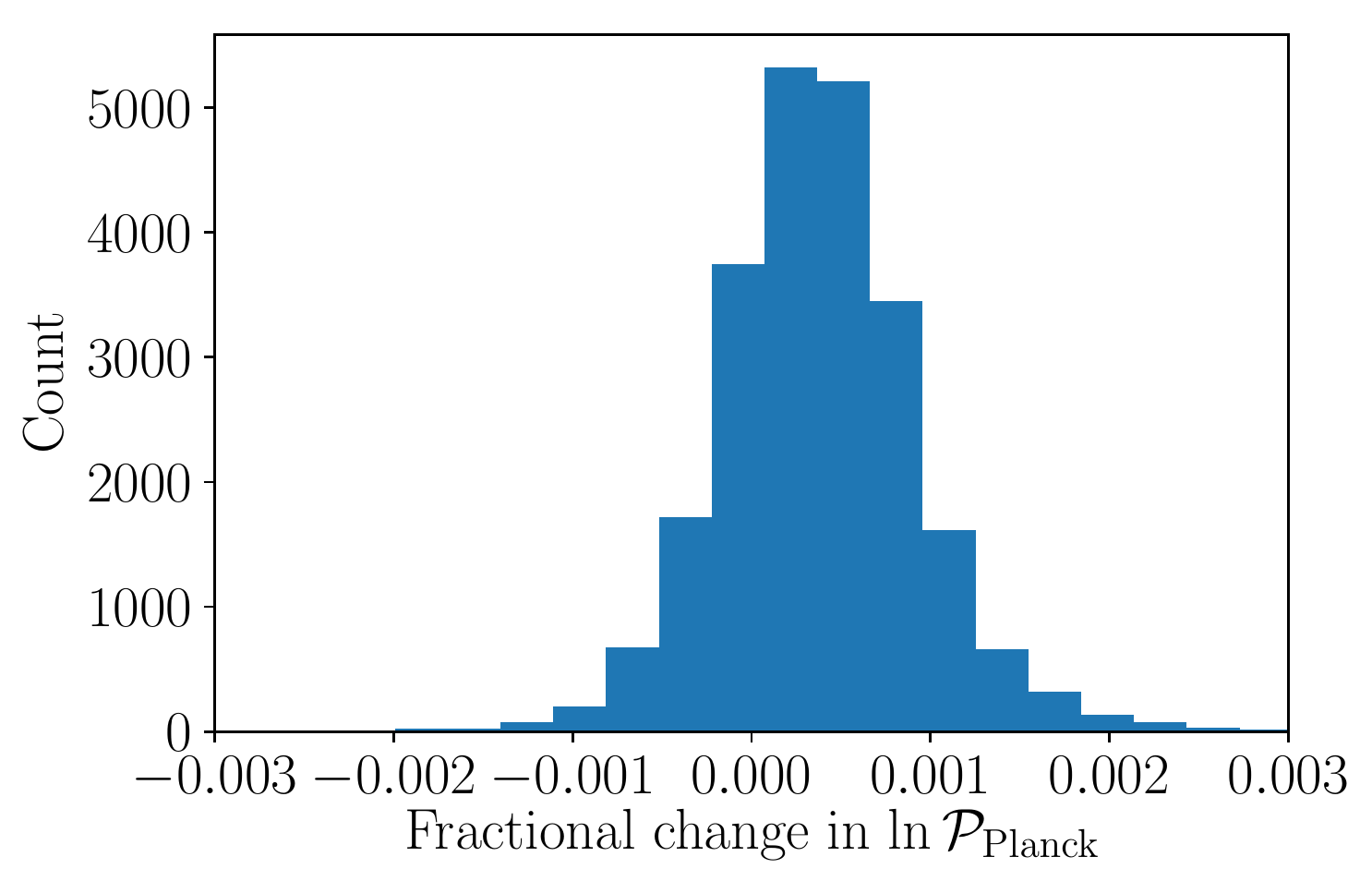}
	\caption{Error in $\ln\mathcal{P}_{\rm Planck}$ for points in the original Planck chain that were not used for reconstructing the posterior probability distribution. Most errors are less than $0.2\%$, meaning our method accurately predicts the $\chi^2$ figure of merit assuming it scales linearly with the log-posterior probability. \label{fig:planck_chi2err}}
	\end{center}
\end{figure}

Resampling our reconstructed posteior yields posteriors that are in excellent agreement with the original chains. Visually inspecting the two distributions appears to show differences in the peaks of the distribution, but this is merely a result of shot noise in the original chains. We verify that our reconstruction accurately recovers the log-posterior probability of the experiment $\ln\mathcal{P}_{\rm Planck}$ as shown in \autoref{fig:planck_chi2err}. This figure shows a histogram of the fractional error in $\ln\mathcal{P}_{\rm Planck}$ predicted by our method for all points in the original MCMC chain that were not used in the reconstruction. The vast majority of points are predicted with better than $0.2\%$ accuracy. Assuming that $\ln\mathcal{P}_{\rm Planck}\propto\chi^2$, this demonstrates that our reconstruction method accurately predicts the $\chi^2$ figure of merit. For reference, we find that the largest shift in the recovered mean and variance among the 6 cosmologically interesting parameters in \planck\ was 0.2\% and 6\% respectively.

To put the power of this new tool into perspective, we note that our reconstructed posterior was created using only 1200 training samples from the original MCMC chain. Sampling from the reconstructed distribution took approximately thirty minutes on a Macbook Pro laptop with a 6-core i7 processor. The speed offered by our method can be used for computationally demanding tasks such as marginalization and computing Bayesian evidences.

\section{Conclusions} \label{sec:conclusions}

We present a method for reconstructing probability distributions given a small, carefully selected sample of points in parameter space at which the posterior is evaluated. This method can be utilized to reconstruct the posterior (or likelihood) function of an experiment given publicly available MCMC data. We demonstrate its utility by showcasing how this technique can be used to jointly sample distributions with partially overlapping parameter spaces. We also showed that we can reconstruct the Planck posterior distribution of the TT, TE, EE+lowE analysis presented in \citet{Planck2018_Cosmology} using only 1200 samples from the original chain. We sample our reconstructed distribution and obtain an MCMC chain with forty times as many samples as the original chain in only thirty minutes.

The utility of being able to reconstruct posteriors from publicly available chains to enable joint analyses was the primary motivator for this work.  However, perhaps the most interesting aspect of the tool we have developed is its potential for significantly reducing the computational effort necessary for generating MCMC samples in the first place.  For instance, while our reconstruction of the \planck\ posterior was based on an existing \planck\ chain, one could have easily adopted a fiducial model in conjunction with a Fisher matrix analysis to determine a few thousand points at which to evaluate the likelihood. The reconstructed posterior can be used to run an MCMC, which can be tested for accuracy through sampling of the Latin Hypercube, and comparing the predicted and actual posterior at the sampled points. In this way, one can determine the regions of parameter space that require further evaluations in order to achieve accurate interpolation. Through iteration, we envision using this method for constructing the full posterior distribution with many fewer likelihood evaluations than brute-force MCMC sampling. This type of approach could be used in conjunction with other proposals such as likelihood-free inference and expectation optimization \citep{Alsing2019_LikeFree,Seljak2019_EL2O}.  We postpone an investigation of the feasibility of such an approach to future work.

\section*{Acknowledgements}

TM thanks V.~Miranda, Niall MacCrann, and An\v{z}e Slosar for helpful discussion.
ER was supported by the DOE grant DE-SC0015975, grant FG-2016-6443, and the Cottrell Scholar program of the Research Corporation for Science Advancement. ER would like to thank Ann Lee for useful conversations that helped spark interest in this work.

\textit{Software}: Python,
Matplotlib \citep{matplotlib},
NumPy \citep{numpy},
SciPy \citep{scipy},
emcee \citep{Foreman13},
george \citep{Ambikasaran2015_George},
ChainConsumer \citep{Hinton2016_ChainConsumer}.

\bibliographystyle{mnras}
\bibliography{main}



\bsp	
\label{lastpage}
\end{document}